\documentclass[letter]{ptptex}

\usepackage{graphicx}
\title{%        %You can use \\ for explicit line-break.
DECIGO/BBO as a probe to constrain \\
alternative theories of gravity%
}

\author{%       %Use \scshape for the family name.
Kent \textsc{Yagi}$^{1}$ %
and Takahiro \textsc{Tanaka}$^{2}$}

\inst{%     %Affiliation, neglected when [addenda] or [errata].
$^1$Department of Physics, Kyoto University,
   Kyoto, 606-8502, Japan \\
$^2$Yukawa Institute for Theoretical Physics,
  Kyoto University,
  Kyoto 606-8502, Japan   
}

\abst{%         %This abstract is neglected when [addenda] or [errata].
We calculate how strongly one can constrain the alternative
 theories of gravity with deci-Hz gravitational wave interferometers
 such as DECIGO and BBO. 
Here we discuss Brans-Dicke theory and massive graviton theories as
 typical examples.
We consider the inspiral of compact binaries
 composed of a neutron star (NS) and an intermediate mass black hole (IMBH)
 for Brans-Dicke (BD) theory and those composed of a super massive black
 hole (SMBH) and a black hole (SMBH) for massive graviton theories. 
Using the restricted 2PN waveforms including spin effects and taking 
the spin precession into account, 
we perform the Monte Carlo simulations of $10^4$ binaries 
to estimate the determination accuracy of binary parameters including
 the Brans-Dicke parameter $\omega_{\mathrm{BD}}$ and the graviton
 Compton length $\lambda_g$. 
Assuming a $(1.4, 10)M_{\odot}$ NS/BH binary of SNR=$\sqrt{200}$, 
the constraint on $\omega_{\mathrm{BD}}$ is obtained as 
$\omega_{\mathrm{BD}}>2.32\times 10^6$, which
is 300 times stronger than the estimated constraint from LISA observation.
Furthermore, we find that, 
due to the expected large merger rate of NS/BH binaries of $O(10^4)$ yr$^{-1}$, 
a statistical analysis yields $\omega_{\mathrm{BD}}>3.77\times10^8$, 
which is 4 orders of magnitude stronger than the current strongest bound obtained from the solar system experiment.
For massive graviton theories, 
assuming a $(10^6, 10^5)M_{\odot}$ BH/BH binary at 3Gpc, one can put 
a constraint $\lambda_g>3.35\times10^{20}$cm, on average. 
This is three orders of magnitude stronger than the one obtained from 
the solar system experiment. 
From these results, it is understood that DECIGO/BBO is a very powerful
 tool for constraining alternative theories of gravity, too.
}

\begin{document}

\maketitle

\quad Many challenges have been made to modify gravitational theory from
general relativity in order to explain the current acceleration of the
universe~\cite{riess}.
In this letter, we focus on two simple possibilities, Brans-Dicke
theory~\cite{brans} and massive graviton
theories~\cite{fierz,rubakov2,dubovsky,rubakov}.
Brans-Dicke theory is a scalar-tensor theory~\cite{fujii} of the
simplest type.
This theory is parameterised by the so-called Brans-Dicke parameter 
$\omega_{\mathrm{BD}}$ and approaches to general relativity in the limit 
$\omega_{\mathrm{BD}}\rightarrow\infty$.
The current strongest bound on this parameter, $\omega_{\mathrm{BD}}>4 \times 10^4$, 
is obtained by the solar system experiment using the Saturn probe satellite Cassini~\cite{cassini}. 
It measured the Shapiro time delay, which corresponds to measuring the spatial metric deviation from general relativity. 
Another constraint on $\omega_{\mathrm{BD}}$ has been obtained from orbital period decay rate of a binary PSR J1141-6545, which consists of a neutron star and a white dwarf~\cite{bhat}.
Since there exists scalar dipole radiation in Brans-Dicke theory, the orbital evolution changes from the one in general relativity, whose information is implemented in the orbital period decay rate.  
Bhat \textit{et al}.~\cite{bhat} found that constraint becomes $\alpha_0^2\equiv \frac{1}{2\omega_{\mathrm{BD}}+3}<2.1 \times 10^{-5}$ for Brans-Dicke theory, which corresponds to $\omega_{\mathrm{BD}} > 2.38 \times 10^4$.
Although this bound is 2 times weaker than the Cassini one, it has distinct meaning since it gives a direct constraint on scalar dipole radiation.
%The constraint from solar system experiment, which is

%Cassini just measured the parameter in the post-Newtonian metric and it was not meauring the effect of scalar dipole radiation.

On the other hand, massive graviton theories by definition introduce a
finite mass $m_g$ to the graviton.
Verification of Kepler's third law in the solar system experiment puts
the lower bound on the graviton Compton length, $\lambda_g\equiv h/m_g
c$, as $\lambda_g>2.8 \times 10^{17}$cm~\cite{talmadge}.

Estimate for the constraint on alternative theories of gravity using
gravitational waves from inspiral compact binaries has been studied in
several papers~\cite{will1994,will1998,scharre,yunes}.
Recently, Berti \textit{et al}.~\cite{berti} calculated the constraint
on $\omega_{\mathrm{BD}}$ and $\lambda_g$ with
LISA~\cite{danzmann,lisa}, using restricted 2PN waveforms and performed
Monte Carlo simulations.
Following their analysis, we improved their calculation by including the
spin-spin coupling $\sigma$, 
small orbital eccentricity, and spin precession effects~\cite{kent}.
Under the assumption of the so-called \textit{simple precession}~\cite{apostolatos,vecchio}
and when we restrict our calculation for circular binary orbits, we obtained the constraints on $\omega_{\mathrm{BD}}$ and $\lambda_g$ as $\omega_{\mathrm{BD}}>6944$ using a $(1.4, 1000)M_{\odot}$ NS/BH binary
of SNR=$\sqrt{200}$ and $\lambda_g>4.86\times10^{21}$cm using a
$(10^7, 10^6)M_{\odot}$ BH/BH binary at 3Gpc.
When we include eccentricities, we found that the constraint on $\omega_{\mathrm{BD}}$ is unaffected as long as we include prior information whilst the one on $\lambda_g$ becomes $\lambda_g>3.10\times10^{21}$cm.
Therefore we can say that the effects of eccentricities are not so strong for both cases.
At the same time, Stavridis and Will~\cite{stavridis} estimated the constraint on $\lambda_g$ for a circular BH/BH binary including both spins of binary objects and taking spin precession into account.
For a BH/BH binary of $(10^6, 10^6)M_{\odot}$ at 3Gpc, they obtained $\lambda_g>5\times10^{21}$cm when the spin-spin precession effect is taken into account and $\lambda_g>4\times10^{21}$cm when it is not taken into account.
To compare our results with their ones, we estimated the constraint on $\lambda_g$ with a $(10^6, 1.1\times 10^6)M_{\odot}$ circular BH/BH binary under simple precession, in which the spin-spin precession effect is neglected, and obtained the constraint $\lambda_g > 3.7\times 10^{21}$cm~\cite{kent}.
Although we cannot directly compare these two results, it seems that our results are consistent with the ones obtained by Stavridis and Will~\cite{stavridis}.
%This proves that our approximation gives a sufficiently correct order of magnitude estimate since we have shown in Table III of Ref.~\citen{kent} that the constraint on $\lambda_g$ with a $(10^6, 10^6)M_{\odot}$ binary do not differ much from the one with a $(10^7, 10^6)M_{\odot}$ binary.
There is also a recent work done by Yunes and Pretorius~\cite{yunes-pretorius} in which they proposed a new framework, \textit{the parametrised post-Einsteinian framework}, to perform gravitational wave tests of alternative theories of gravity. 

Following our previous paper~\cite{kent}, we estimate the possible
constraint on $\omega_{\mathrm{BD}}$ and $\lambda_g$ obtained by 
detecting gravitational waves from the inspiral of precessing compact
binaries using deci-Hz space-borne gravitational wave interferometers
such as DECIGO~\cite{seto,kawamura,sato} and
BBO~\cite{phinneybbo,ungarelli}.
These detectors are most sensitive in the frequency band between 0.1Hz
and 1Hz, and the noise levels are about four orders of magnitude lower
than that of LISA.
These detectors have a huge number of compact 
binaries as promising sources. (Because of that, high precision cosmology 
using them is also expected~\cite{cutlerholz}.)
In this letter we perform the analysis assuming DECIGO noise curve but
almost the same results will apply to BBO, too. 
Here, we restrict our attention to circular binaries since the lower bounds on
$\omega_{\mathrm{BD}}$ and $\lambda_g$ are not much affected by the inclusion of
eccentricity into binary parameters.
(The upper bound on $\omega_{\mathrm{BD}}$, if
detected, can be affected by including eccentricity, though.)
Since the detection rate of NS/BH mergers is expected to be $O(10^4)$ for DECIGO/BBO, we add a statistical analysis, which improves the constraint on the deviation from general relativity.
Our results are only approximate estimates in that we do not take the errors coming from
the use of approximate waveforms into account~\cite{cutler-vallisneri} and also due to the limitation of the validity of the Fisher analysis~\cite{vallisneri}. 

%For Brans-Dicke theory, we will obtain the constraint
%$\omega_{\mathrm{BD}}>1.676\times 10^6$ for a $(1.4+10)M_{\odot}$ NS/BH
%binary of SNR=$\sqrt{200}$.
%This is 200 times stronger than the one obtained in Ref.~\cite{kent}
%using LISA.
%For massive graviton theories, we will obtain $\lambda_g>2.24\times10^{20}$cm for a $(10^6+10^5)M_{\odot}$ BH/BH binary at 3Gpc.
%This is one order of magnitude worse than the one obtained in
%Ref.~\cite{kent} but still three orders of magnitude better than the
%one obtained from the solar system experiment.
%See Ref.~\cite{kent} for more details.

%The remaining of this letter is organised as follows.
%In Sec.~\ref{sec-decigo}, we review the basic plan of DECIGO and show
%the noise curve.
%In Sec.~\ref{sec-results}, we show the results of our calculations.
%In Sec.~\ref{sec-conc},  we conclude this letter with some topics left
%for future works.   
%We take $c=G=1$ and cosmological parameters as $\Omega_m=0.3$,
%$\Omega_{\Lambda}=0.7$ and $H_0=72$km/s/Mpc.

%\section{\label{sec-decigo}DECIGO}
 
%\vspace{3mm}

\quad First we review the basic plan of DECIGO (DECi-hertz Interferometer Gravitational wave Observatory) and show the noise curve.
DECIGO is a planned space gravitational wave antenna mission~\cite{seto,sato}.
It consists of four constellations of three drag free satellites which keep the triangular shape throughout the flight.
These three satellites form Fabry-Perot interferometers with separation 1000km.
DECIGO can detect gravitational waves from both astrophysical and cosmological sources mainly between 0.1-10Hz.
Among them, compact binaries are the promising sources, although the
primary goal of the mission is to detect primordial gravitational wave
background at the level of $\Omega_{\mathrm{GW}}=10^{-16}$.
For LISA frequency band, such small signals are completely obscured by the foregrounds generated by white-dwarf binaries.
Since this foreground noise has cut off frequency around 0.2Hz~\cite{farmer}, DECIGO has a much better chance to detect the primordial gravitational wave background.

%Each satellite contains a 100kg mirror with 1m diameter.
%The laser power is 10W with wavelength $\lambda=532$nm.

The instrumental noise spectral density for DECIGO is given by~\cite{ando} 
\begin{equation*}
S_h^{\mathrm{inst}}(f)=5.3\times 10^{-48}
\Biggl[(1+x^2)+\frac{2.3\times 10^{-7}}{x^4(1+x^2)}
+\frac{2.6\times 10^{-8}}{ x^4}\Biggr]\mathrm{Hz^{-1}},
\end{equation*}
where $x=f/f_p$ with $f_p\equiv 7.36 \mathrm{Hz}$. 
The three terms on the right hand side represent the shot noise, the radiation pressure noise and the acceleration noise, respectively. 

Besides instrumental noise, there are three main foreground confusion
noises: galactic white dwarf binaries
$S_h^{\mathrm{gal}}(f)$~\cite{nelemans}, extra-galactic white dwarf
binaries $S_h^{\mathrm{ex-gal}}(f)$~\cite{farmer}, and neutron star
binaries $S_h^{\mathrm{NS}}(f)$~\cite{cutler-harms}.
The noise spectral densities for the first two confusion noises are given in Ref.~\citen{kent}. 
Following Ref.~\citen{cutler-harms}, the noise spectral density for NS binaries is estimated as
$
S_h^{\mathrm{NS}}(f)=1.3\times 10^{-48} \left( f/1\mathrm{Hz} \right)^{-7/3},
$
where we have set the cosmological parameters to $\Omega_m=0.3$, $\Omega_{\Lambda}=0.7$ and $H_0=72$km/s/Mpc.
 
\begin{figure}[thbp]
  \centerline{\includegraphics[scale=.5,clip]{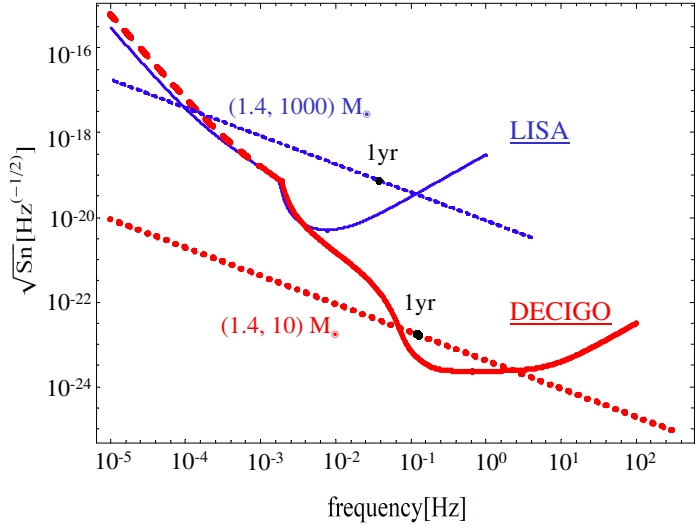} }
 \caption{\label{noise-decigo}
 The noise spectral density for DECIGO (thick solid curve) and LISA (thin solid curve).
It might be possible to extend the DECIGO noise curve down to 10$^{-5}$Hz~\cite{ando} (thick dashed curve). 
We also show the amplitudes of gravitational waves from a $(1.4, 1000)M_{\odot}$ (thin dotted line) and a $(1.4, 10)M_{\odot}$ (thick dotted line) NS/BH binary with each SNR set to $\rho =10$ for 1yr observation before coalescence. 
Each dot labeled ``1yr'' represents the frequency at 1yr before coalescence. }
\end{figure}

Then, the total noise spectral density for DECIGO becomes
%\begin{widetext}
%\begin{eqnarray}
%S_h(f)&=&\min\Bigl[ \frac{S_h^{\mathrm{inst}}(f)}{\exp(-k
% N_f/T)},
% S_h^{\mathrm{inst}}(f)
%+S_h^{\mathrm{gal}}(f){\cal R}(f)\Bigr]
%\cr&&
%+S_h^{\mathrm{ex-gal}}(f){\cal R}(f)
%+0.01\times S_h^{\mathrm{NS}}(f), \label{noise-DECIGO}
%\end{eqnarray}

\begin{equation}
S_h(f)=\min\Bigl[ \frac{S_h^{\mathrm{inst}}(f)}{\exp(-k
 N_f/T)},
 S_h^{\mathrm{inst}}(f)
+S_h^{\mathrm{gal}}(f){\cal R}(f)\Bigr]
+S_h^{\mathrm{ex-gal}}(f){\cal R}(f)
+0.01\times S_h^{\mathrm{NS}}(f), \label{noise-DECIGO}
\end{equation}
%\end{widetext}
where the factor 
${\cal R}(f)\equiv \exp\{-2\left(
	{f}/{0.05\mathrm{Hz}} \right)^2\}$ 
represents the cutoff of the WD/WD binary confusion noise.
We put the factor 0.01 in front of $S_h^{\mathrm{NS}}(f)$, 
which represents the fraction of gravitational waves that 
cannot be removed after foreground subtraction. 
Namely, we assume that 99$\%$ of the gravitational waves from neutron
star binaries can be identified and removed.
$N_f$ is the number density of galactic white dwarf binaries per unit
frequency given in Ref.~\citen{kent}.
$k\simeq 4.5$ is the average number of frequency bins that are lost when each galactic binary is fitted out and $T$ represents the observation time which we fix as 1yr.
The noise curve (\ref{noise-DECIGO}) is shown in Fig.~\ref{noise-decigo} as a thick solid curve.
We introduce low and high cut-off frequencies at 
$f_{\mathrm{low}}=10^{-3}$Hz and $f_{\mathrm{high}}=100$Hz, respectively. 
Although it has not been estimated rigorously, it might be possible to extend the noise curve on lower frequency side down to $10^{-4}$-$10^{-5}$Hz~\cite{ando}.
This bound comes from the limitation of controlling the mirror positions.
This extension is shown as a thick dashed curve in Fig.~\ref{noise-decigo}.
We also show the noise curve of LISA as a thin solid curve for comparison. 
Thin dotted line and thick dotted line each, respectively, represent the amplitude of gravitational waves from a $(1.4, 1000)M_{\odot}$ and a $(1.4, 10)M_{\odot}$ NS/BH binary of SNR $\rho=10$ with 1yr observation before coalescence.
Each dot labeled ``1yr'' represents the frequency at 1yr before coalescence.

%\section{\label{sec-results}NUMERICAL CALCULATIONS AND RESULTS}
%\vspace{3mm}

We used the matched filtering analysis to estimate the determination accuracy of the binary parameters. %~\cite{finn,flanagan}. 
There are 14 parameters in total:
the chirp mass $\ln\mathcal{M}$, the dimensionless mass ratio $\eta$;
the coalescence time $t_c$, the coalescence phase $\phi_c$;
the distance to the source $D$;
the spin-orbit and spin-spin coupling coefficients, $\beta$ and $\sigma$;
the directional cosine between the orbital and spin angular momenta $\kappa$, 
the precession angle parameter $\alpha_c$; 
the direction of the source $(\theta_{\mathrm{S}},\phi_{\mathrm{S}})$; 
the initial direction of the total angular momentum
$(\theta_{\mathrm{J}},\phi_{\mathrm{J}})$; 
finally, $\omega_{\rm BD}^{-1}$ or $\lambda_g^{-1}$, 
depending on which theory we are aiming to constrain.
We calculate the inverse of the Fisher matrix numerically, assuming stationary and
Gaussian noise.
Integration is performed from $f_{\mathrm{in}}$ to $f_{\mathrm{fin}}$, where $f_{\mathrm{in}}=\max \bigl\{ f_{\mathrm{low}},  f_{1\mathrm{yr}} \bigr\}$ and $f_{\mathrm{fin}}=\min \bigl\{ f_{\mathrm{high}},  f_{\mathrm{ISCO}} \bigr\}$, respectively.
$f_{1\mathrm{yr}}$ is the frequency at 1yr before coalescence and $f_{\mathrm{ISCO}}$ is the one at the innermost stable circular orbit (ISCO). 
First we performed the pattern-averaged estimate of the errors in 
determination of the parameters for binaries with various masses, in which we have
averaged over the directions of the source and the orbital angular
momentum, and the spins of the binary constituents are assumed to be zero. 
We set SNR=10 for Brans-Dicke theory and $D_L=3$ Gpc for massive graviton theory.
We also performed the Monte Carlo simulations 
for $10^4$ binaries, % and take the average at the end.
distributing 
$(\theta_{\mathrm{S}},\phi_{\mathrm{S}},\theta_{\mathrm{J}},\phi_{\mathrm{J}},\kappa,\alpha_c)$
randomly, 
both with and without the precession effect. 
We set the fiducial values to 
$t_c=\phi_c=\omega_{\rm BD}^{-1}=\lambda_{g}^{-1}=0$ with
$(m_{\mathrm{NS}},m_{\mathrm{BH}})=(1.4,10)M_{\odot}$ of SNR=10 (for
pattern-averaged estimate) or SNR=$\sqrt{200}$ (for Monte Carlo
simulations) for Brans-Dicke theory, and
$(m_{\mathrm{BH}},m_{\mathrm{BH}})=(10^6,10^5)M_{\odot}$ at $D_L=3$Gpc
for massive graviton theories. 
For the analysis
without the spin precession effect, 
the fiducial values of $\beta$ and $\sigma$ are set to 0. 
When we take it into account, we set the dimensionless spin parameter
$\chi\equiv S/m^2$ to 0 and 0.5 for the lighter and heavier bodies of
binaries, respectively, where $S$ is the magnitude of 
the spin angular momentum. 
We include prior information on $\beta$ and $\sigma$ when calculating Fisher matrices. 
See Ref.~\citen{kent} for more details.

\begin{table}[t]
\caption{\label{table-bd-noangle} The results of error estimation with DECIGO in
 Brans-Dicke theory for various mass NS/BH binaries. We performed 
pattern-averaged estimates, using only one
 detector and fixing the SNR to 10.}
%\begin{ruledtabular}
\begin{center}
\begin{tabular}{c||ccccccc}  %\hline
 masses & $f_{\mathrm{in}}$ & $f_{\mathrm{fin}}$ & $\omega_{\mathrm{BD}}$ &  $\Delta \ln\mathcal{M}$ & $\Delta\ln \eta$ & $\Delta \beta $ & $\Delta \sigma$  \\ 
 &  (Hz) &  (Hz) & $(10^6)$ & $(10^{-5})$ & $(\%)$ & &  \\ \hline
% &  & ($\%$) & & & & ($10^{-3}$str) & \\ \hline\hline 
%\multicolumn{6}{l}{(1.4+400)$M_{\odot}$} \\
$(1.4, 10)M_{\odot}$ & 0.118 & 100 & 1.342 & 0.978 & 2.78 & 0.190  & 2.18 \\
$(1.4, 50)M_{\odot}$ & 0.0776 & 85.6 & 0.2662 &  2.34 & 2.64 & 0.106 & 1.09 \\ 
$(1.4, 100)M_{\odot}$ & 0.0651 & 43.36 & 0.1899 & 2.34 & 1.87 & 0.0485 & 0.563  \\
$(1.4, 400)M_{\odot}$ & 0.0460 & 10.95 & 0.04244 & 4.96 & 1.85 & 0.0133 & 0.250  \\
\end{tabular}
\end{center}
%\end{ruledtabular}
\end{table}

\begin{table}[t]
\caption{\label{table-bd-compare} Comparison of the constraints on $\omega_{\mathrm{BD}}$ and other physical quantities for a $(1.4, 10)M_{\odot}$ binary with DECIGO and a $(1.4, 1000)M_{\odot}$ binary with LISA. We performed pattern-averaged analyses and SNRs are fixed to 10 for both cases.}
%\begin{ruledtabular}
\begin{center}
\begin{tabular}{c||cccccc}  %\hline
 masses and detector & $\omega_{\mathrm{BD}}$ &  $\omega_{\mathrm{BD}}^{\mathrm{uncor}}$ & $N_{\mathrm{GW}}$ & $v_{\mathrm{1yr}}$ & $f_{\mathrm{in}}$ & $f_{\mathrm{fin}}$ \\ 
 &  ($10^6$) & ($10^6$) & $(10^6)$ &  & (Hz) & (Hz)  \\ \hline
$(1.4, 10)M_{\odot}$, DECIGO & 1.34 & 332 & 5.9 & 0.027 & 0.118 & 100   \\
$(1.4, 1000)M_{\odot}$, LISA & 0.00821 & 21.6 & 1.8 &  0.083 & 0.0366 & 1.00 \\ 
\end{tabular}
\end{center}
%\end{ruledtabular}
\end{table}

In Table~\ref{table-bd-noangle}, we show the pattern-averaged results of binary parameter estimation errors for Brans-Dicke theory with $(1.4,10)M_{\odot}$, $(1.4,50)M_{\odot}$, $(1.4,100)M_{\odot}$ and $(1.4,400)M_{\odot}$ NS/BH binaries of SNR=10.
(It seems that SNRs of $O(10)$ are too small in performing the Fisher analysis.~\cite{vallisneri}
However, constraints for higher SNR binaries are obtained by just scaling in proportional to SNR. )
It can be seen that smaller mass binaries give stronger constraints on $\omega_{\mathrm{BD}}$.
This can be understood as follows.
The velocities of binaries at 1 yr before coalescences are slower for smaller mass binaries.
Since Brans-Dicke theory gives dipole correction to binary gravitational waves, this correction is -1PN order.
Therefore, when we fix SNRs, this contribution is larger for slower binaries, which makes the constraints stronger. 
Comparing these results with the ones in Ref.~\citen{kent}, we can say that DECIGO has better ability in constraining $\omega_{\mathrm{BD}}$ compared to LISA.
To be more explcit, let us compare the constraint from $(1.4, 10)M_{\odot}$ with DECIGO and the one from $(1.4, 1000)M_{\odot}$ with LISA.
This mass parameter is an optimised choice for each detector to constrain Brans-Dicke theory.
First, we compare the uncorrelated constraint $\omega_{\mathrm{BD}}^{\mathrm{uncor}}$ which is calculated directly from the Fisher matrix (not the inverse of it) and represents the possible constraint when the degeneracies between $\omega_{\mathrm{BD}}$ and other parameters have been solved completely~\cite{berti}.
%We found that $\omega_{\mathrm{BD}}^{\mathrm{uncor}}$ for $(1.4, 10)M_{\odot}$ with DECIGO and for $(1.4, 1000)M_{\odot}$ with LISA are 3.32$\times$ $10^8$ and 2.16$\times$ $10^7$, respectively.
Table~\ref{table-bd-compare} shows that the former constraint is stronger than the latter by more than 1 order of magnitude.
There are 2 reasons for this, (1) the number of gravitational wave cycles $N_{\mathrm{GW}}$ are greater and (2) the velocity at 1 yr before coalescence $v_{\mathrm{1yr}}$ is slower.
%We estimated $N_{\mathrm{GW}}$ for $(1.4, 10)M_{\odot}$ binary to be 5.9$\times$ $10^6$ whereas the one for $(1.4, 1000)M_{\odot}$ binary is 1.8$\times$ $10^6$.
%Therefore the former $N_{\mathrm{GW}}$ is about 3 times greater than the latter.
%Also, $v_{\mathrm{1yr}}$ for $(1.4, 10)M_{\odot}$ binary and $(1.4, 1000)M_{\odot}$ binary is estimated as 0.027 and 0.083, respectively.
From Table~\ref{table-bd-compare}, we can see that $N_{\mathrm{GW}}$ for the former case is about 3 times greater compared to the latter case.  
On the other hand, $v_{\mathrm{1yr}}$ for the former one is 3 times slower than the latter. 
Since the -1PN correction term in the phase is proportional to $v^{-2}$, this contribution is about 10 times larger for the former case.
These 2 contributions make the former constraint greater by more than 1 order of magnitude.
Next, we compare the constraint on $\omega_{\mathrm{BD}}$ which is calculated from the inverse of the Fisher matrices.
%A $(1.4, 10)M_{\odot}$ binary with DECIGO gives $\omega_{\mathrm{BD}} > 1.34 \times 10^6$ whilst a $(1.4, 1000)M_{\odot}$ binary with LISA gives  $\omega_{\mathrm{BD}} > 8.21 \times 10^3$.
From the table, we understand that the correlation between other parameters for the former case is weaker by more than 1 order of magnitude.
We think that this comes from the difference in the width of effective frequency range of observation.
From the table, we see that this frequency range is larger for the former case by more than 1 order of magnitude, which solves the degenaracies between parameters.

In Table~\ref{table-massive-noangle}, we show the pattern-averaged results for massive graviton theory with $(10^6,10^6)M_{\odot}$, $(10^6,10^5)M_{\odot}$, $(10^5,10^5)M_{\odot}$, $(10^5,10^4)M_{\odot}$ and $(10^4,10^4)M_{\odot}$ BH/BH binaries.
In this case, larger mass binaries give stronger constraints on $\lambda_g$.
This is explained from the correction to the phase velocity $v_{ph}$ which is given as~\cite{kent}
\begin{equation}
v_{\mathrm{ph}}^2=\left( 1-\frac{1}{f^2\lambda_g^2}\right)^{-1}.
\end{equation}  
Since more massive binaries generate lower frequency gravitational waves, these binaries give larger corrections and make the constraints stronger.
However, if we increase the mass parameter too much, the SNR starts to decrease since the dynamical frequency range starts to get narrower, which makes the constraints weaker.
This is why the constraint on $\lambda_g$ from a $(10^6,10^6)M_{\odot}$ binary is weaker compared to the one from a $(10^6,10^5)M_{\odot}$ binary.  
In Ref.~\citen{kent}, we estimated the constraint from a $(10^7,10^6)M_{\odot}$ binary with LISA to be $4.06\times 10^{20}$cm.
In this case LISA gives stronger constraint on $\lambda_g$ than DECIGO.
This is because LISA is able to detect lower frequency gravitational waves than DECIGO where the correction on $v_{ph}$ is greater.
If we use the extended version of DECIGO shown as thick dashed curve in Fig.~\ref{noise-decigo}, we found that $(10^7,10^6)M_{\odot}$ binary gives a constraint of $4.05\times 10^{20}$cm which coincides with the one obtained with LISA.
This result is obvious since the frequency range of this binary is $f=2.36\times 10^{-5}-4.00\times 10^{-4}$Hz in our case and the noise curves of extended DECIGO and LISA are almost identical within this frequency range.

\begin{table}[t]
\caption{\label{table-massive-noangle} The results of error estimation in
 massive graviton theory for BH/BH binaries at 3 Gpc with various masses. We performed 
pattern-averaged estimates using only one
 detector.}
%\begin{ruledtabular}
\begin{center}
\begin{tabular}{c||cccccccc}  %\hline
 masses & $f_{\mathrm{in}}$ & $f_{\mathrm{fin}}$ & SNR & $\lambda_g$ &  $\Delta \ln\mathcal{M}$ & $\Delta\ln \eta$ & $\Delta \beta $ & $\Delta \sigma$  \\ 
 &  (mHz) &  (mHz) & & $(10^{20}\mathrm{cm})$ & $(\%)$ &  & & \\ \hline
% &  & ($\%$) & & & & ($10^{-3}$str) & \\ \hline\hline 
%\multicolumn{6}{l}{(1.4+400)$M_{\odot}$} \\
$(10^6, 10^6)M_{\odot}$ & 1.0 & 2.20 & 1338 & 1.014 & 14.0 & 2.46 & 9.40 & 2.50 \\
$(10^6, 10^5)M_{\odot}$ & 1.0 & 4.00 & 2044 & 1.270 & 1.19 & 1.69 & 9.10 & 2.44 \\ 
$(10^5, 10^5)M_{\odot}$ & 1.0 & 22.0 & 4909 & 1.133 & 0.0286 & 0.930 & 7.00 & 1.69  \\
$(10^5, 10^4)M_{\odot}$ & 1.0 & 40.0 & 3021 & 0.4066 & 4.51$\times 10^{-3}$ & 0.823 & 6.20 & 1.88  \\
$(10^4, 10^4)M_{\odot}$ & 1.0 & 220.0 & 29569 & 0.3852 & 3.54 $\times 10^{-4}$ & 0.924 & 6.96 & 1.68  \\
\end{tabular}
\end{center}
%\end{ruledtabular}
\end{table}

 \begin{table}[t]
\caption{\label{table-bd-noprec} The results of error estimation in
 Brans-Dicke theory for $(1.4, 10)M_{\odot}$ NS/BH binaries. The first
 line shows the results of pattern-averaged estimate. We used only one
 detector and the SNR is fixed to 10. The second and the third lines
 show the results of Monte Carlo simulations. We used two detectors for
 the analyses and we set SNR$=\sqrt{200}$ (corresponding to SNR=10 for
 each detector). We distribute $10^4$ binaries, calculate the
 determination error of each parameter for each binary and take the
 average. The second line shows the ones without taking spin precession
 into account, whilst the third line represents the ones including
 precession. $\sigma$ is included in the binary parameters for all the
 cases.}
%\begin{ruledtabular}
\begin{center}
\begin{tabular}{c||ccccccc}  %\hline
 cases & $\omega_{\mathrm{BD}}$ &  $\Delta \ln\mathcal{M}$  & 
                    $\Delta\ln \eta$ & $\Delta \beta $ &
                    $\Delta \ln D_L$ & $\Delta \Omega_S$ & $\Delta \sigma$  \\ 
 & $(10^6)$ & $(10^{-5})$ & $(\%)$ & & & $(10^{-5}\mathrm{str})$ & \\ \hline
% &  & ($\%$) & & & & ($10^{-3}$str) & \\ \hline\hline 
%\multicolumn{6}{l}{(1.4+400)$M_{\odot}$} \\
pattern-averaged & 1.342 & 0.978 & 2.78 & 0.190 & 0.100 & - & 2.18 \\
no precession & 0.9774 &  1.22 & 3.06 & 0.186 & 1.24 & 3.27 & 2.15   \\ 
including precession & 2.317 & 0.350 & 0.295 & 0.0551 & 0.183 & 2.52 & 0.627  \\
\end{tabular}
\end{center}
%\end{ruledtabular}
\end{table}

\begin{figure}[h]
  \centerline{\includegraphics[scale=.4,clip]{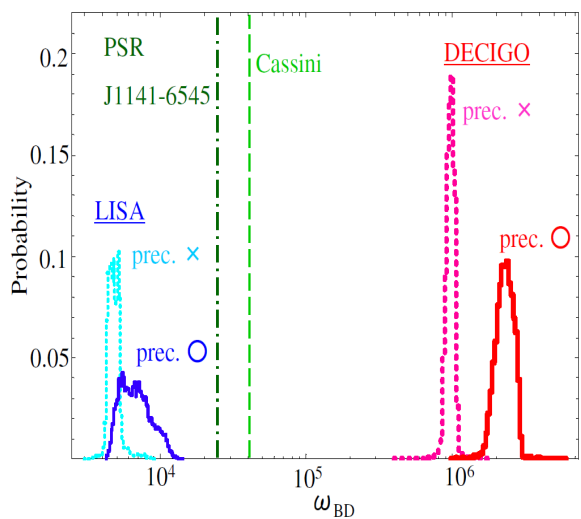} }
 \caption{\label{bd} The histograms showing the probability distribution
 of the lower bound of $\omega_{\mathrm{BD}}$ obtained from our Monte
 Carlo simulations of $10^4$ NS/BH binaries in Brans-Dicke theory. 
We take the masses of the binaries as $(1.4, 10)M_{\odot}$ with
 SNR$=\sqrt{200}$. 
The thick dotted one represents the estimate without precession and
 the thick solid one shows the one including precession using
 DECIGO. For comparison, we also show the results obtained
 in~\cite{kent} for $(1.4, 1000)M_{\odot}$ NS/BH binaries with SNR$=\sqrt{200}$ using LISA. 
The thin dotted one shows the one without precession and the
 thin solid one represents the one including precession. The dashed line at $\omega_{\mathrm{BD}}=2.38\times 10^4$ represents the constraint from PSR J1141-6545~\cite{bhat}.}
\end{figure}

In Table~\ref{table-bd-noprec}, we show the results of error estimation
in Brans-Dicke theory for $(1.4, 10)M_{\odot}$ NS/BH binaries with both pattern-averaged analysis and Monte Carlo simulations.
%of
%SNR=10 or $\sqrt{200}$.
%\textbf{Here, we have chosen an optimised mass parameter which can be seen from Table~\ref{table-bd-noangle}.}
The first row shows the ones with pattern-averaged estimate where we used only one detector and fixed the SNR to 10.
The second and the third rows represent the results of Monte Carlo simulations with SNR fixed to $\sqrt{200}$.
The second row shows the ones without taking spin precession into account whilst the third row shows the ones including precession.
We see that inclusion of precession improves the constraint on $\omega_{\mathrm{BD}}$ by a factor of two. 
In general, binary parameters including $\omega_{\mathrm{BD}}$ are correlated with spin parameters $\beta$ and $\sigma$.
When we include precession, we obtain additional information about spin which solves degeneracies and reduces the estimation errors, making the constraint stronger.
Figure~\ref{bd} represents the histograms showing the number fraction of binaries which give the constraint on the Brans-Dicke parameter within each bin of $\omega_{\mathrm {BD}}$. 
The thick dotted one represents the estimate without precession and the thick solid one shows the one including precession using DECIGO. 
For comparison, we also show the results obtained in~\cite{kent} for $(1.4, 1000)M_{\odot}$ NS/BH binaries with SNR=$\sqrt{200}$ using LISA. 
The thin dotted one shows the one without precession and the thin solid one represents the one including precession.
The dashed line at $\omega_{\mathrm{BD}}=1.47\times 10^5$ represents the constraint from PSR J1141-6545~\cite{bhat}.
We can see that DECIGO can put 300 times stronger constraint than LISA.
%This is mainly because the number of cycles of gravitational wave
%$N_{\mathrm{GW}}\equiv\int^{f_{\mathrm{fin}}}_{f_{\mathrm{in}}}df
%(f/\dot{f})$ is larger, and also because the noise levels of DECIGO are
%lower than that of LISA.
The reasons are the same as the pattern-averaged analysis.

Unlike the case of LISA, these binaries are thought to be the definite sources for DECIGO.
The event rate of NS/NS binary mergers is estimated to be
$10^5$ yr$^{-1}$~\cite{cutler-harms},
and the rate of NS/BH mergers will be about one order of magnitude
smaller than 
that of NS/NS mergers (see Shibata \textit{et al}.~\cite{shibata} and references therein).
Therefore it is possible to put even stronger constraint by performing a
statistical analysis.
Defining the variance of parameter
$\bar{\omega}=\omega_{\mathrm{BD}}^{-1}$ from $i$-th binary as
$\sigma_i$, the total variance $\sigma$ is given by 
\begin{equation}
\sigma^{-2}%=\sum_i \sigma^{-2}_i
=\Delta T \int^{\infty}_{0} 4\pi [a_0 r(z) ]^2
\dot{n}(z)\frac{d\tau}{dz}\sigma (z)^{-2} dz, 
\label{statistics}
\end{equation}
where $\Delta T =1$ yr represents the observation time, $a_0$ represents
the current scale factor, $r(z)$ is the comoving distance to the source,
$\dot{n}(z)$ shows the NS/BH merger rate at redshift $z$ and $\tau$ is
the proper look back time of the source.
$a_0r(z)$ and $\frac{d\tau}{dz}$ are given as~\cite{cutler-harms}
$a_0r(z) ={H_0^{-1}}\int ^z_0 {dz'}/\sqrt{\Omega_m (1+z')^3+\Omega_{\Lambda}}$, 
${d\tau}/{dz} = \{{H_0}(1+z){\sqrt{\Omega_m (1+z')^3+\Omega_{\Lambda}}}\}^{-1}, 
$
respectively.
Following Ref.~\citen{cutler-harms}, we adopt the model for $\dot{n}(z)$ 
given by $\dot{n}(z)=\dot{n}_0 R(z)$, 
where $\dot{n}_0=10^{-8}$ Mpc$^{-3}$ yr$^{-1}$ is the estimated 
merger rate today and 
$R(z)= 1+2z  ~(\rm{for}~z\leq 1),~ 
           (3/4)(5-z) ~(\rm{for}~1\leq z\leq 5),~
           0 ~(\rm{for}~ z\geq 5)
$
encodes the time-evolution of this rate.
This model gives the merger rate of $10^4$ yr$^{-1}$.
For simplicity, we assume that all NS/BH binaries have the same typical masses of $(1.4, 10)M_{\odot}$. 
We first calculate the variance $\sigma (z)$ for various $z$ 
using pattern-averaged estimate and obtain the total variance 
$\sigma$ from Eq.~(\ref{statistics}).
%Then we calibrate the result by taking into account the difference 
%between the Monte Carlo analysis and the pattern-averaged estimate.
The Fourier component of the pattern-averaged waveform is given by 
Eq.~(29) of Ref.~\citen{kent}.
To take into account the effects of redshift, all we have to do is to
replace the masses with the redshifted ones: 
$m_{\mathrm{NS}}\rightarrow (1+z)m_{\mathrm{NS}}$
and $m_{\mathrm{BH}}\rightarrow (1+z)m_{\mathrm{BH}}$. 
The distance in this expression is to be understood as the 
luminosity distance given by 
$D_L=a_0 (1+z)r(z)$.
%To calculate $\sigma$ using Eq.~(\ref{statistics}), we equally divide the redshift ranging from $z=0$ to $z=5$ into 1000 bins.
%Then the integral in Eq.~(\ref{statistics}) can be reformulated into summation as
%\begin{equation}
%\sigma^{-2}=4\pi \Delta T \sum_{i=1}^{1000}[a_0 r(z_i) ]^2 \dot{n}(z_i)\frac{d\tau}{dz}\Big |_{z=z_i}\sigma (z_i)^{-2} \Delta z,
%\end{equation}
%where $z_i\equiv (i-1/2)\Delta z$ and $\Delta z \equiv 5/1000$.
%We calculate $\sigma (z_i)$ for each $z=z_i$ using pattern-averaged waveform including $z$ dependence.

From the pattern-averaged analysis, we find that observation of 
$10^4$ binaries with $(1.4, 10)M_{\odot}$ can put a constraint 
$\omega_{\mathrm{BD}}>2.18\times 10^8$.
This is 94 times stronger than the one from a single binary 
placed at 17 Gpc (corresponding to a binary of SNR=10).
We calibrate the result of this analysis by using the results of Monte Carlo simulation 
to yield $\omega_{\mathrm{BD}}>3.77\times 10^8$.
This is 4 orders of magnitude stronger than the current strongest bound.  
%The distance, out to which the NS/BH merger rate becomes 1/yr, is 3000Mpc.
%In this case, the constraint becomes $\omega_{\mathrm{BD}}>7.66\times 10^6$ for $(1.4+10)M_{\odot}$ binaries.
%This is 200 times stronger than the Cassini bound.

\begin{table}[t]
\caption{\label{table-massive-noprec} The results of error estimation in
 massive graviton theories for $(10^6, 10^5)M_{\odot}$ BH/BH binaries at
 3Gpc. The meaning of each line is the same as in
 Table~\ref{table-bd-noprec}.}
%\begin{ruledtabular}
\begin{center}
\begin{tabular}{c||ccccccc}  %\hline
 cases & SNR & $\lambda_g$ & $\Delta \ln\mathcal{M}$  & 
                    $\Delta\ln \eta$ & $\Delta \beta $ & $\Delta \sigma$ &
                     $\Delta \Omega_S$   \\ 
 & & $(10^{20}\mathrm{cm})$  & $(\%)$ & &  & & (str) \\ \hline
% & ($10^{21}$cm) & & ($\%$) & & & & ($10^{-4}$str) & \\ \hline\hline 
pattern-averaged & 2044 & 1.270  & 1.19 & 1.69 & 9.10 & 2.44 & - \\ 
no precession & 2601  & 1.266 & 1.16 & 1.64 & 8.92 &  2.40 & 1.16  \\ 
including precession & 2666 & 3.349 & 0.314 & 0.0388 & 0.0612 & 0.529 &  0.0248 \\ %\hline
\end{tabular}
\end{center}
%\end{ruledtabular}
\end{table}

\begin{figure}[htbp]
  \centerline{\includegraphics[scale=.4,clip]{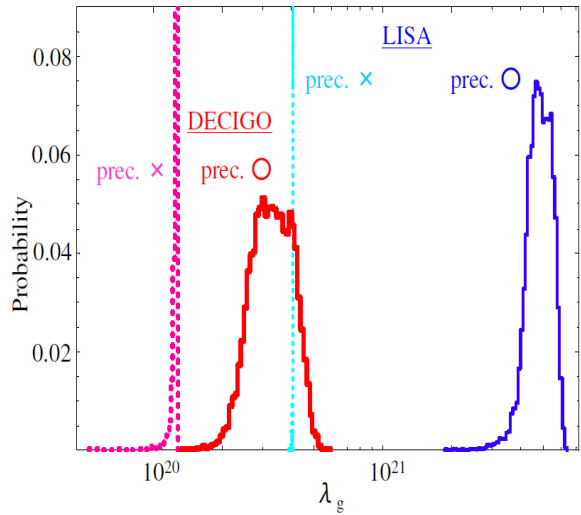} }
 \caption{\label{massive} The histograms showing the probability
 distribution of the lower bound of $\lambda_g$ obtained from our Monte
 Carlo simulations of $10^4$ BH/BH binaries in massive graviton
 theories. We take the masses of the binaries as $(10^6, 10^5)M_{\odot}$
 at 3Gpc. For comparison, we also show the results obtained
 in~\cite{kent} for $(10^7, 10^6)M_{\odot}$ BH/BH binaries using
 LISA. The meaning of each histogram is the same as in Fig.~\ref{bd}.}
\end{figure}

In Table~\ref{table-massive-noprec}, we show the results of error estimation in massive graviton theories for $(10^6, 10^5)M_{\odot}$ BH/BH binaries at 3Gpc with both pattern-averaged analysis and Monte Carlo simulations..
Again, we have chosen an optimised mass parameter which can be understood from Table~\ref{table-massive-noangle}.
The meaning of each row is the same as in Table~\ref{table-bd-noprec}.
We can see that the constraint on $\lambda_g$ increases by a factor of two when we include precession.
Figure~\ref{massive} represents the histograms showing the number fraction of binaries that give the constraint of each $\lambda_g$. 
For comparison, we also show the results obtained in~\cite{kent} for $(10^7, 10^6)M_{\odot}$ BH/BH binaries at 3 Gpc using LISA.
SNRs for the gravitational waves from these binaries correspond to 1600.
The meaning of each histogram is the same as in Fig.~\ref{bd}.
We can see that the effect of precession is larger for LISA.
We expect that this is because the effective frequency range is wider for $(10^7, 10^6)M_{\odot}$ BH/BH binaries with LISA. 
The lower bound on $\lambda_g$ obtained by DECIGO, $3.35\times 10^{20}$cm, is 
one order of magnitude smaller than that obtained by LISA.
%This is because the masses of binaries are smaller for DECIGO.
However, this is still three orders of magnitude larger than 
the one obtained by the solar system experiment.
These results show how powerful DECIGO can be in constraining alternative theories of gravity.

Recently, Arun and Will~\cite{arun-will} have estimated the constraint
on $\lambda_g$ including higher harmonics in the waveform.
Since they do not include spins, it is important to include both higher
harmonics and spin precession, and estimate the constraint on various
alternative theories of gravity.
This is left for future work.  
%\begin{acknowledgments}

We thank Naoki Seto, Takashi Nakamura and Masaki Ando for useful discussions and
numerical code corrections, and Bernard Schutz for valuable comments.
This work is in part supported by the Grant-in-Aid for Scientific
Research Nos. 19540285 and 21244033.
This work is also supported in part by the Grant-in-Aid for the Global
COE Program ``The Next Generation of Physics, Spun from Universality and
Emergence'' from the Ministry of Education, Culture, Sports, Science and
Technology (MEXT) of Japan.

%\section*{Acknowledgements}
%We would like to thank ...........

%\appendix
%\section{First Appendix} %Empty argument \section{} yields `Appendix'. 
%
%\section{Second Appendix}


\begin{thebibliography}{99}

\bibitem{riess}
Supernova Search Team Collab. (A. G. Riess \textit{et al.}), Astrophys. J. \textbf{607}, 665 (2004).

\bibitem{brans}
C. Brans and R. H. Dicke, Phys. Rev. \textbf{124}, 925 (1961). 

\bibitem{fierz}
M. Fierz and W. Pauli, Proc. Roy. Soc. Lond. \textbf{A173}, 211 (1939). 

\bibitem{rubakov2}
V. A. Rubakov, hep-th/0407104. 

\bibitem{dubovsky}
S. L. Dubovsky, JHEP \textbf{0410}, 076 (2004).

\bibitem{rubakov}
V. A. Rubakov and P. G. Tinyakov, Phys. Usp. \textbf{51}, 759 (2008). 

\bibitem{fujii} e.g. 
Y. Fujii and K. Maeda, \textit{The Scalar-Tensor Theory of Gravitation}, Cambridge University Press (2007).

%\bibitem{comment}
%Although we mentioned that the current strongest constraint on $\omega_\mathrm{BD}$ was the Cassini bound in Ref.~\cite{kent}, we later found that the one from PSR J1141-6545 was greater than the Cassini bound.

\bibitem{cassini}
B. Bertotti, L. Iess and P. Tortora, Nature \textbf{425}, 374 (2003). 

\bibitem{bhat}
N. D. R. Bhat, M. Bailes and J. P.W. Verbiest, Phys. Rev. \textbf{D77}, 124017 (2008). 

\bibitem{talmadge}
C. Talmadge, J. P. Berthias, R. W. Hellings and E. M. Standish, Phys. Rev. Lett. \textbf{61}, 1159 (1988). 

\bibitem{will1994}
C. M. Will, Phys. Rev. \textbf{D50}, 6058 (1994). 

\bibitem{will1998}
C. M. Will, Phys. Rev. \textbf{D57}, 2061 (1998). 

\bibitem{scharre}
P. D. Scharre and C. M. Will, Phys. Rev. \textbf{D65}, 042002 (2002). 

\bibitem{yunes}
C. M. Will, and N. Yunes, Class. Quant. Grav. \textbf{21}, 4367 (2004).

\bibitem{berti}
E. Berti, A. Buonanno and C. M. Will, Phys. Rev. \textbf{D71}, 084025 (2005). 

\bibitem{danzmann}
K. Danzmann, Class. Quant. Grav. \textbf{14}, 1399 (1997). 

\bibitem{lisa}
LISA web page: http://lisa.nasa.gov/

\bibitem{kent}
K. Yagi and T. Tanaka, Phys. Rev. \textbf{D81}, 064008 (2010).

\bibitem{apostolatos}
T. A. Apostolatos, C. Cutler, G. J. Sussman, and K. S. Thorne, Phys. Rev. \textbf{D49}, 6274 (1994). 

\bibitem{vecchio}
A. Vecchio, Phys. Rev. \textbf{D70}, 042001(2004). 

\bibitem{stavridis}
A. Stavridis, C. M. Will, arXiv:0906.3602 [gr-qc]. 

\bibitem{yunes-pretorius}
N. Yunes and F. Pretorius, Phys. Rev. \textbf{D80}, 122003 (2009). 

\bibitem{seto}
N. Seto, S. Kawamura and T. Nakamura, Phys. Rev. Lett. \textbf{87}, 221103 (2001). 

\bibitem{kawamura}
S. Kawamura \textit{et al}., Class.Quant.Grav. \textbf{23}, S125 (2006). 

\bibitem{sato}
S. Sato \textit{et al}., J. Phys. Conf. Ser. \textbf{154}, 012040 (2009).

\bibitem{phinneybbo} 
E. S. Phinney \textit{et al}., \textit{Big Bang Observer Mission Concept Study} (NASA), (2003).

\bibitem{ungarelli}
C. Ungarelli, P. Corasaniti, R. A. Mercer and A. Vecchio, Class. Quant. Grav. \textbf{22}, S955 (2005). 

\bibitem{cutlerholz}
C. Cutler and D. E. Holz, arXiv:0906.3752 [astro-ph.CO]. 

\bibitem{cutler-vallisneri}
C. Cutler and M. Vallisneri, Phys. Rev. \textbf{D76}, 104018 (2007). 

\bibitem{vallisneri}
M. Vallisneri, Phys. Rev. \textbf{D77}, 042001 (2008). 

\bibitem{farmer}
A. J. Farmer and E. S. Phinney, Mon. Not. Roy. Astron. Soc. \textbf{346}, 1197 (2003). 

%\bibitem{noise}
%private communication

\bibitem{ando}
M.~Ando, private communications.

\bibitem{nelemans}
G. Nelemans, L. R. Yungelson, and S. F. Portegies Zwart, Astron and Astrophys. \textbf{375}, 890 (2001).

\bibitem{cutler-harms}
C. Cutler and J. Harms, Phys. Rev. \textbf{D73}, 042001 (2006). 

%\bibitem{phinney}
%E. S. Phinney, astro-ph/0108028.

%\bibitem{allen}
%B. Allen, gr-qc/9604033.

%\bibitem{finn}
%L. S. Finn, Phys. Rev. \textbf{D46}, 5236 (1992). 

%\bibitem{flanagan}
%C. Cutler and E. E. Flanagan, Phys. Rev. \textbf{D49}, 2658 (1994). 

\bibitem{shibata}
M. Shibata, K. Kyutoku, T. Yamamoto and K. Taniguchi, Phys. Rev. \textbf{D79}, 044030 (2009). 

\bibitem{arun-will}
K. G. Arun and C. M. Will, arXiv:0904.1190 [gr-qc]. 


%%%%%%%%%%%%%%%%%%%%%%%%%%%%%%%%%%%%%%%%%%%%%%%%%%%%%%%%%%%%%
% Some macros are available for the bibliography:
%  o for general use
%    \JL : general journals                 \andvol : Vol (Year) Page
%  o for individual journal 
%    \AJ   : Astrophys. J.           \NC         : Nuovo Cim.
%    \ANN  : Ann. of Phys.           \NPA, \NPB  : Nucl. Phys. [A,B]
%    \CMP  : Commun. Math. Phys.     \PLA, \PLB  : Phys. Lett. [A,B]
%    \IJMP : Int. J. Mod. Phys.      \PRA - \PRE : Phys. Rev. [A-E]     
%    \JHEP : J. High Energy Phys.    \PRL        : Phys. Rev. Lett.
%    \JMP  : J. Math. Phys.          \PRP        : Phys. Rep.
%    \JP   : J. of Phys.             \PTP        : Prog. Theor. Phys.     
%    \JPSJ : J. Phys. Soc. Jpn.      \PTPS       : Prog. Theor. Phys. Suppl.
% Usage:
%  \PRD{45,1990,345}          ==> Phys.~Rev.\ D \textbf{45} (1990), 345
%  \JL{Nature,418,2002,123}   ==> Nature \textbf{418} (2002), 123
%  \andvol{123,1995,1020}    ==> \textbf{123} (1995), 1020
%%%%%%%%%%%%%%%%%%%%%%%%%%%%%%%%%%%%%%%%%%%%%%%%%%%%%%%%%%%%%
  


\end{thebibliography}
\end{document}